# An Intelligent Data Analysis for Hotel Recommendation Systems using Machine Learning

**Bushra Ramzan, Imran Sarwar Bajwa, Noreen Jamil, Farhaan Mirza**

School of Computer Science & IT, University of Birmingham, B15 2TT, United Kingdom.
*Corresponding author: i.s.bajwa@cs.bham.ac.uk;

**Abstract**: This paper presents an intelligent approach to handle heterogeneous and large-sized data using machine learning to generate true recommendations for the future customers. The Collaborative Filtering (CF) approach is one of the most popular techniques of the RS to generate recommendations. We have proposed a novel CF recommendation approach in which opinion based sentiment analysis is used to achieve hotel feature matrix by polarity identification. Our approach combines lexical analysis, syntax analysis and semantic analysis to understand sentiment towards hotel features and the profiling of guest type (solo, family, couple etc). The proposed system recommends hotels based on the hotel features and guest type as additional information for personalized recommendation. The developed system not only has the ability to handle heterogeneous data using big data Hadoop platform but it also recommend hotel class based on guest type using fuzzy rules. Different experiments are performed over the real world dataset obtained from two hotel websites. Moreover, the values of precision and recall and F-measure have been calculated and results are discussed in terms of improved accuracy and response time, significantly better than the traditional approaches.

---

## 1. Introduction

With the evolution of new web technologies, the recommender systems (RS) are getting significant attention by the business people as well as customers due to its role in better e-commerce, refined business strategy, improved customer's satisfaction, etc. The success of modern e-commerce systems and online booking and reservations systems heavily relies on the customer's satisfaction and trust. Mariani et al. [1] has observed that tourism is one of the most famous and a powerful industry in the world which has a huge impact on the world's total GDP or employment. Tourism is linked with hotels because tourists always wanted to know about the hotels where they are going to stay in their tour. In the recent years, online hotel booking has become one of the primary choices of the hotel customers. A few recommender systems are also developed in the recent past to facilitate the hotel customers to recommend a hotel before he actually makes a booking or a reservation Liu et al. [2]. However, these systems are generic and process only homogenous data; whereas nature of most of the data on web is heterogeneous that is a major bottleneck in the performance of hotel recommendation systems. The heterogeneity of data affects the performance of the RS directly. In this era of competition, complex information causes overload problems which in turn are time consuming and affects the overall performance. Due to the various forms of data (numeric, textual, etc) in heterogeneous form over the web, the performance of RSs requires more attention towards its improvement.

Few recommenders dealing heterogeneous data are now available in market in recent by Li et al. [44]. Our proposed recommender system uses a Hotel feature utility matrix to recommend a suitable hotel to a user on the basis of both quantitative (numerical) and qualitative (textual) features to achieve true recommendations. Secondly, a fuzzy module provides the recommendation of hotels in a particular type of user's such as solo, family, business, friends, couple etc because recommendations will be different based on type of user trip and user preference. Like for a family, "room" and "food" and "cleanliness" facilities are the main preferences but for a single guest, facilities like "pool", "spa" and "gym" may have a greater preference. Similarly "Wifi and "computer" can be an important feature for the user who is on the Business trip. A true recommendation targets a relevant recommendation with respect to a customer's choices, priorities, budget, etc. Typically, a recommender system banks on the previous information such as customers' reviews and ratings about the hotel's attributes or features Zhang et al. [3]. A typical subset of a dataset is shown in Figure 1 containing hotel ratings and reviews. A couple of challenges in achieving true hotel recommendations is processing and analysis of large heterogeneous web data and intelligent approach that makes recommendations relevant to customer's choice.

{"ratings": {"service": 5.0, "cleanliness": 5.0, "overall": 5.0, "value": 5.0, "location": 5.0, "sleep_quality": 5.0, "rooms": 5.0}, "title": "\u201cMy home away from home!\u201d", "text": "On every visit to NYC, the Hotel Beacon is the place we love to stay. So conveniently located to Central Park, Lincoln Center and great local restaurants. The rooms are lovely - beds so comfortable, a great little kitchen and new wizz bang coffee maker. The staff are so accommodating and just love walking across the street to the Fairway supermarket with every imaginable goodies to eat (if you choose not to go out for every meal!)", "author": {"username": "Maureen V", "num_reviews": 2, "num_cities": 2, "id": "E3C85CA9DBBBC77E0DB534ABE93E4713", "location": "Sydney, New South Wales, Australia"}, "date_stayed": "December 2012", "offering_id": 93338, "num helpful votes": 0, "date": "December 17, 2012", "id": 147639004, "via mobile": false}

**Figure 1.** Subset of a hotel dataset (numerical & textual)

While dealing with diverse nature of data in our multi feature hotel recommendation system, the main challenge was the opinion mining/sentiment analysis of users' reviews to calculate a polarity score which represents the degree of likeness or dis-likeness about a hotel by a user. This polarity score provides a textual feature of user's opinions about a particular hotel. To handle the diversity of heterogeneous data as the presented approach uses both numeric data as well as textual data, a big data solution involving Hadoop was used in our approach because it efficiently handles data heterogeneity and data diversity in a better way. We have defined the guest type (solo, family, business, friends, couple) as the main part of this research. We will not only consider different rating parameters but also apply feature based sentimental analysis on user's reviews. For example, TripAdvisor allows travelers to rate hotels on several options such as such as location, room, cleanliness, service and staff. The process of extracting opinions from textual reviews is called as Sentiment analysis/opinion mining.

The target of our recommender system is to record preferences and choices of customers and to make suggestions or recommendations according to one's choices and preferences. Koren et al. [4] has discussed that the quality of recommendations are also considerably important when large amount of user ratings and reviews needed to be processed to provide efficient and true recommendations . In recent years, booking of hotels through online systems has been increasing rapidly and many websites are working over this domain. A recommendation system (RS) helps customers in effective ways to get information about different products. A generic recommender architecture is presented in Figure 2. Hsieh et al. [5] has discussed that different algorithms and techniques are used in recommender system for considering the attribute of the users' reviews and ratings. R. Burke [6] has explain that such algorithms are divided into collaborative filtering (CF) and content-based filtering (CBF). The mixer of these algorithms is known as hybrid filtering.

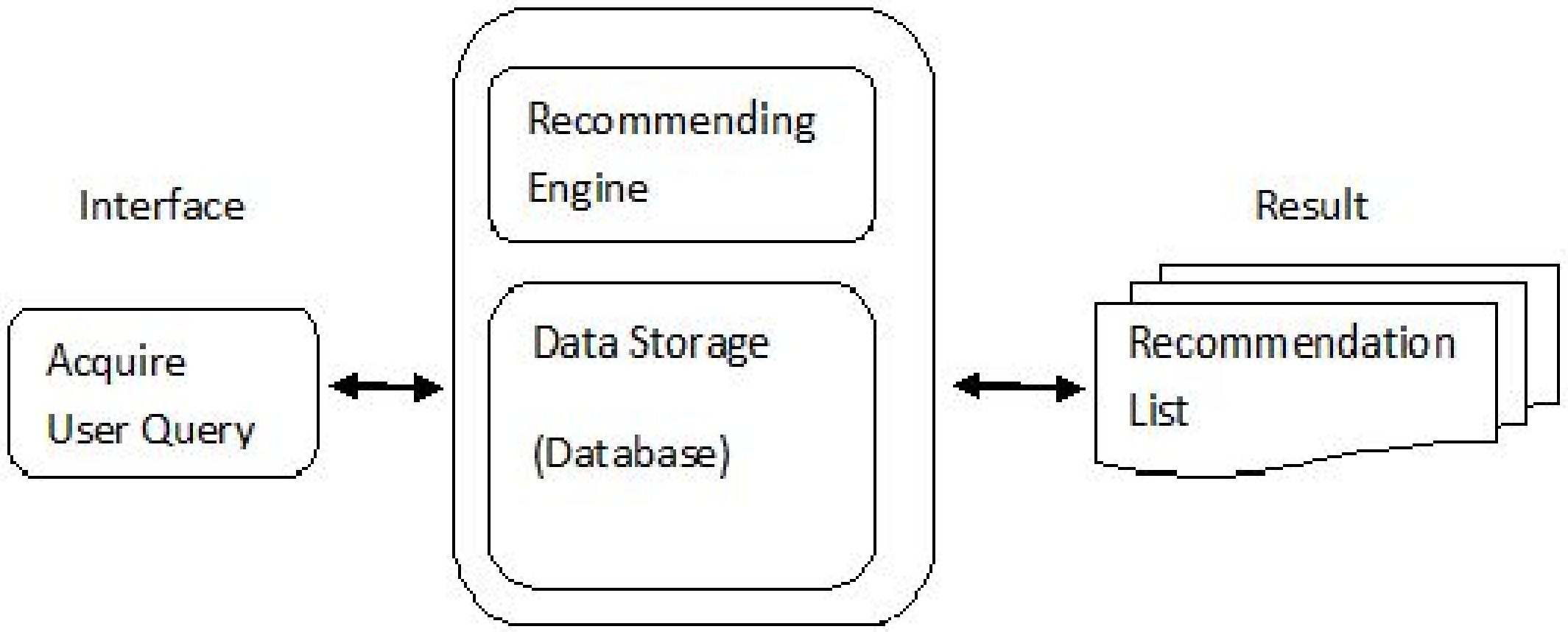

**Figure 2.** Generic architecture of a recommender system

Lops et al. [7] elaborated that, In a recommender system with content-based, the user's preferences are presented by their linked points. On the other hand the collaborative filtering is the most extensive recommender technique. CF works by collecting user ratings for items in a given domain and calculating similarities between users or items in order to provide relevant recommendations. Ekstrand et al. [8] discuss in detail about collaborative filtering and explain that it is a recommended technique and works as a class of methods that recommend items to users based on the preferences other users have expressed for those items. It deals with different types of data i.e. hotels, movies, music where the user preferences are changed randomly.

There are two kinds of recommender system based on Collaborative filtering algorithms such as item-based recommender system and user based recommender system. Item-based recommenders differ from the user based recommenders only because they compare item similarities instead of user similarities.

According to Zhang et al. [9] Collaborative Filtering (CF) based approach is very successful technology in all RSs. The [10],[11] has reported that there are three fundamental challenges faced by CF approaches such as

- **Cold start** challenge rise where an item appears which has not been rated before, recommendations cannot be made for it or when a new user without any prerecorded profile appears [12, 13].
- **Sparsity** challenge appears when there are numerous items but too less rating values are available in the initial stage of recommendation [10],[13].
- **Scalability** challenge appears when users' and items' data is very big to process [10].

Some researchers have introduced another approach known cluster based approach [5, 10, 14]. Collaborative filtering based on Clustering reduces computation time and focuses only on time efficiency improvement as the clustering phase is performed off-line.

The main idea here is to develop a recommender system which helps users to find hotels according to their preference and choice using previous users' reviews and ratings. The core problem in Recommenders is storage and processing of the list of thousands of items like hotels here. To resolve this issue a recommendation system is proposed here which is based on item-based collaborative filtering using Hadoop containing NoSQL database in a fault-tolerant and concurrent manner for the sake of improving the performance and efficiency due to the huge size of data of the hotels and users expeditious reviews. We have tested the performance gains of using Hadoop in terms of response time of the web application. It also resolves the scalability issue of collaborative filtering Recommenders because it is difficult for a system to generate recommendations in practice if the users and items databases are very huge and require massive computation [10]

The proposed system is suitable to store and retrieve the required information efficiently and compute high speed recommendations within seconds. This paper aimed to make the following main contributions

- The proposed recommender system helps in achieving true hotel recommendations through processing and analyzing of large heterogeneous web data using opinion mining approach and fuzzy approach to produce relevant recommendations according to customers' type and choice.
- Our proposed recommender system uses a Hotel feature utility matrix to recommend a list of suitable hotels to a user on the basis of both ratings (numerical) and reviews (textual) to achieve true recommendations.
- A fuzzy module provides the recommendation of hotels in a particular type of Guest type such as solo, family, business, friends', couple etc.
- Development of a web based application for hotel recommendation by incorporating hotel information from the linked data of external resources (websites) available online.
- The proposed approach optimizes performance in the proposed hotel recommendation system using NoSQL Cassandra database in Hadoop Environment.
- Dataset is obtained from two different sources (websites) such as TripAdvisor.com, Expedia.com.

We are using opinion based approach which is used to classify the text into three sentiment expression such as "Positive", "Negative and Neutral with the help of SentiWordnet Wordnet dictionaries. There were number of challenges that arise during the processing of reviews and extraction of the features from textual reviews. Some of the challenges are given below.

- Dealing with the big data which consist of the textual reviews describing opinions given by the people for the hotels.
- Casual informal languages, abbreviation/emoji/slang or use of emoticons.
- Spelling mistakes/ typing errors.
- Ambiguous reviews given by a customer. E.g. I have never lived in a hotel quite like this one before! Ambiguity: we cannot understand; either the hotel is best or worst
- Reviews containing hashtags.
- Detecting polarities of hidden sentiments of a customer in a given review.

The rest of the paper is organized and structured as follows. Section 2 provides the detailed description and explanation of related work and associated concepts. Collaborative filtering is discussed to build a recommender system. A General model of a recommender system is explained for the understanding of our proposed recommender application. Section 3 describes the methodology used to design a hotel recommender application. It gives the specifics of the designed hotel recommender components and interface. Section 4 provides the details of preliminary experiments and results along with the system overview. In section 5 a detailed comparative analysis is performed upon the existing studies with the proposed approach. In the end, section 6 provides conclusion and future directions related to our work. It also presents the implementation details and the main features of the developed system.

## 2. Related Work

The initial focus of the paper is to analyze and understand prior research and work done in recommender system concepts such as information processing, analysis applications, and recommendation service [15]. In this part of the paper we have studied and analyzed the literature related to existing approaches which are similar to the proposed idea. These related studies have become an inspiration to understand and solve problems which help to realize the recommender systems' impact on consumer behavior and decisions.

*2.1. Recommender Systems*

Recommender systems are the systems which help consumers discover items they may like. Fasahte et al. [16] have concluded that many researchers have conducted several experiments on the TripAdvisor dataset. Different filtering technique has been used to predict the unrated items. In addition to these approaches, authors also proposed a hybrid approach in recommendation systems to analyze the customer behavior by using rating data of the customer review and textual content as well. Tan et al. [12] has worked on a novel similarity measure inspired by a physical resonance phenomenon, named resonance similarity (RES) is proposed proving superior predictive accuracy as compared to the existing similarity measures on users' evaluations. Crespo et al. [45] discuss that Sem-Fit uses the customers' experience point of view in order to apply fuzzy logic methods to relating customer and hotel characteristics, represented by means of domain ontologies and affect grids. Hu et al. [17] explain that the data for the analysis containing reviews and ratings is obtained from TripAdvisor. The performance of the Context Aware Personalized Hotel (CAPH) recommendation system is evaluated by measuring the fineness of rating predictions.

The Hwang et al. [19] focus on B&BS reviews and hotels to perform a hotel review for the hotel management systems. Hotel reviews were obtained from TripAdvisor.com. Latent Dirichlet Allocation (LDA) method was used to obtain the comparable performance to the Term Frequency-Inverse Document Frequency (TF-IDF) method through the high recall and with fewer features. The author has concluded that all types of features used in the analysis are beneficial to predict the reviews of a noteworthy hotel. An unexpected result is that the semantic-based LDA method possesses less precision as compared to the word based LDA method. The Sandeep et al. [46] performed twitter community sentiment analysis to obtain real time sentiments of the common people to represent both existing and potential customers. The proposed method using month wise sentiment score of twitter hash tags of Indian telecom operators successfully predicted their growth rate in terms of subscriber addition. A method called KASR (Keyword Aware Service Recommendation) has been proposed for the big data analysis of reviews by Meng et al. [18]. In a big data environment, KASR has been implemented in Hadoop and cloud for the sake of improving the efficiency and the scalability in the environment of big data[20, 21].

Lin et al. [22] designed a personalized hotel recommendation by using the text browsing tracking and mining techniques. Rianthong et al. [23] developed a useful stochastic programming model to plan the hotels sequence in order to enable the customers to search the hotels at a lower search cost. Multiple regression analysis was used to test the data. Findings show that the hotels with review rating, higher utility and prices should be kept at the upper part of the sequence. Lal et al. [47] explores most relevant and crucial features for sentiment classification and groups them into seven categories, named as, Basic features, Seed word features, TF-IDF, Punctuation based features, Sentence based features, N-grams, and POS lexicons. Liu et al. [1] introduced the idea for the accurate recommendation by combining the opinions and preferences of the users. Sharma et al. [24] has examined a recommendation system by using a multi-criteria review-based approach based on the user's reviews and preferences. A dataset from a website Booking.com was used for the analysis. The NLP was used for the sake of determining the rating of the hotel used by the previous customers.

Chang et al. [25] hypothesized on the recommendation of the hotel that is based on the surrounding environments. Real-world check-in spots dataset are used to check the recommendations of the proposed framework. Various analytical approaches for reviews such as CATPAC (Content analysis program) software were used to analyze the consumers' reviews and ratings while SPSS (Statistical Package for Social Sciences) was used to analyze the customers' ratings. Regression analysis and Analysis of Variance (ANOVA) support the findings to know about the images of hotel brands to check the customer loyalty. Jannach et al. [26] has worked out on the recommendation of hotels that is based on the multi-dimensional customer ratings for Movie lens movies. Authors utilized the regression-based methods and user-and- item-based models for accurate and efficient recommendations.

A model of the personalized Intelligent Information for the hotel services is examined by the Chawla et al. [27]. An investigation performed on the recommendations for improving the news articles through the user clustering is performed by Bouras et al [28]. Word Net-enabled k-means algorithm is used. Chen et al. [29] examined a nonlinear and Fuzzy programming approach for the optimization of the performance of ubiquitous hotel recommendation system using hotels dataset. Fasahte et al. [16] has explored the hotel recommendation system using Lexicon based approach to search sentiment exposure towards the hotel's aspects relied on the defined context. Secondly, the Item-based CF technique has been used to predict the unrated items. A hybrid recommendation approach to analyze the user's behavior by using both eating data of the user's review, textual content and rating data of user's is used. Valcarce et al. [30] analyzed the platform of the distributed recommendation for the big data using MySQL Cluster and Cassandra.

Collaborative filtering-systems gather user's previous data for the certain items like hotels, books, Articles etc. A hotel recommendation system based on Rankboost algorithm and cluster based collaborative filtering is proposed by Huming et al. [10] to help users to choose a hotel according to their desire. The data for the experiment collected from hoteltravel.com. For the quantitative and qualitative analysis, the 5 point rating scale was used.

Several researchers have applied collaborative filtering algorithms [31-37] for recommendations. Some of them are discussed below. Moreover, in order to calculate the correlation with other users or to do deductions in the feature space collaborative and content-based filtering depends on knowledge of the user [38, 39]. The current proposed hotel recommender system is highly trusted and somewhat bridge the hotel feature extraction using Natural language opinion mining analysis.

A multi-criteria collaborative filtering for the tourism domain by using PCA-ANFIS (Principal Component Analysis-Adaptive Neuro-Fuzzy Inference System) and EM (Expectation Maximization) is implemented on a Recommender system [1]by Nilashi et al. [40]. Several experiments have been done on the dataset obtained from TripAdvisor website. EM, ANFIS and PCA methodology are used for the analysis which achieved the high accuracy for the recommendation of the hotel and improved the time efficiency. Phorasim et al. [41]. investigated on the recommender system for the movies through the use of k-means and collaborative filtering as well. It is concluded that the proposed algorithm is more precise and less time consuming as compared to the existing traditional one. K. Stefan [42] suggests that in order to infer list of relevant and actionable suggestions in real time, data from different sources can be combined by the recommender system for model driven software engineering. Do et al. [43] surveys common techniques for implementing model-based approach so as to achieve high accuracy Yibo et al. [48] has built a hybrid recommendation model for movie

1

recommendation using sentiment analysis on spark platform outperforms the traditional models in terms of various evaluation criteria.

**3. Hotel Recommender System**

The proposed system uses the heterogeneous nature of data (textual, numerical) crawled in from World Wide Web (www). Data is obtained from the selected hotel websites (data sources) containing the keywords present in the active user search query. A web crawler was used to download the requested data and store the obtained data in a NoSQL database cassandra for further processing. The data is usually found in the form of numeric (such as votes, ranks and number of video views) and text (such as reviews and comments). To get true recommendations our system has used ranks, votes and reviews data to extract Hotel features from it.

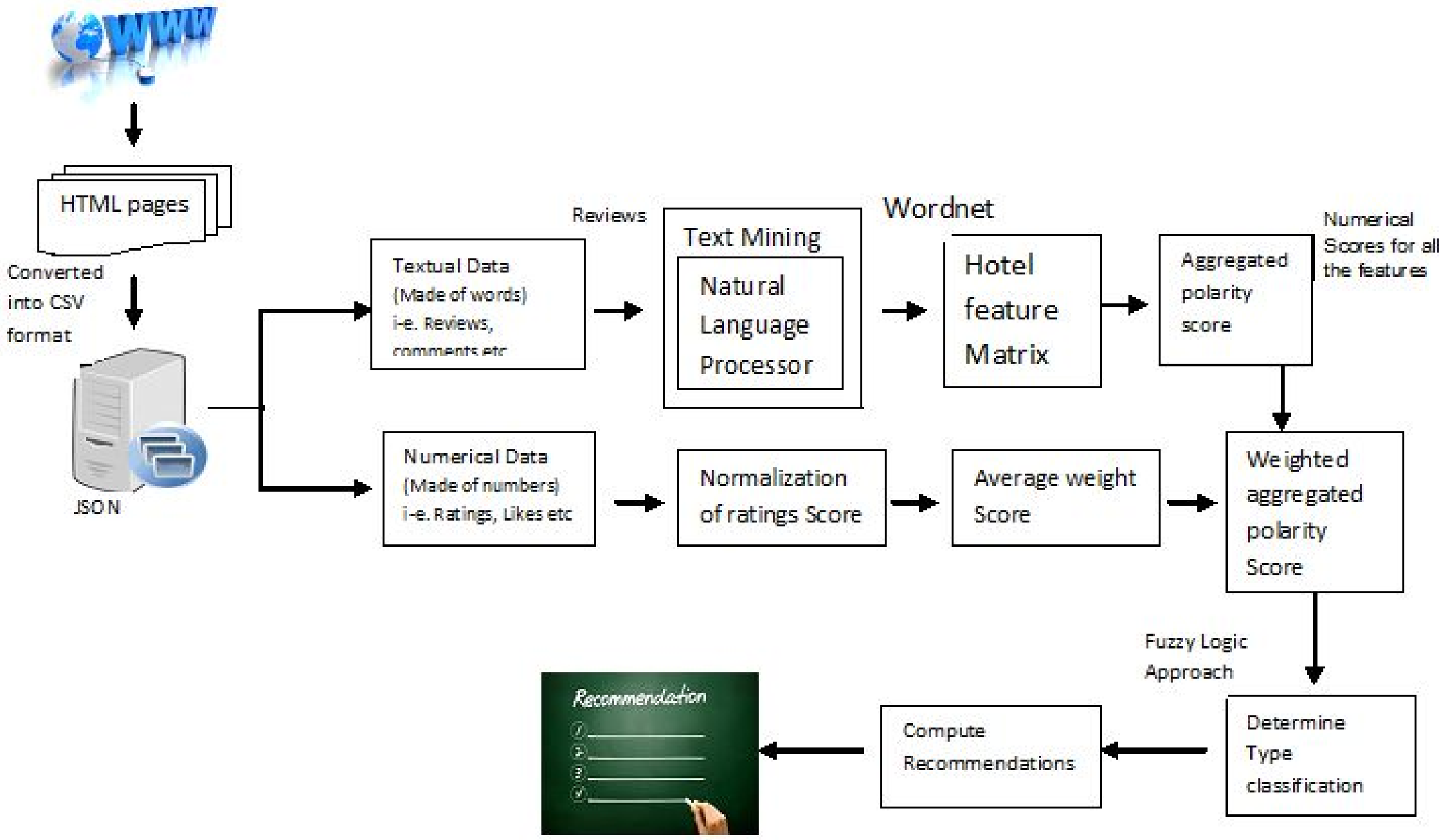

**Figure 3.** The proposed approach for hotel recommendation

The system works in two parallel ways. The numeric ranks and votes of hotels from each selected data source are normalized. On the other hand Review data is processed using Natural language processing package for review mining and features are extracted in the form of a hotel feature matrix. Further, numerical polarity scores are computed for these extracted features using SentiWordNet and average polarity score is calculated. Now weighted average polarity scores are calculated by aggregating normalized rank score, voting score and polarity score. Finally, recommendations are computed by applying the fuzzy logic approach. We have defined a fuzzy set containing certain fuzzy rules to calculate the final score to find out the guest type (solo, family, business, friends, couple etc.) for the hotel. The proposed Hotel recommendation approach is shown in Figure 3. The final recommendations of the Hotels based on a particular guest type in one of the five different classes are displayed.

*3.1. Feature Extraction Process*

First of all, before deriving reviewer's feature preferences, we first analyze raw textual reviews and convert them into structured form to extract opinion based feature. Reviews from any website are usually extracted into a Json file which is then loaded into the system database.

We have studied different types of methods for mining the feature based opinions from textual reviews, and found that NLTK package is most appropriate tool to extract features from hotel reviews. Natural language toolkit NLTK is a Python library to make programs that work with natural language. The library can perform different operations such as tokenizing, stemming, classification, parsing, tagging, and semantic reasoning. We have used NLTK 3.3**.** Version in this paper. Following steps have been performed for identifying numerous features from the hotel reviews such as:

- Extracting features from a review and grouping synonymous features
- Findings and assigning value to the opinions that are associated with various features in the review
- Assigning these features a value in the normalized range.

The review data is converted in comma separated values to be available in easily readable form i-e Natural language data format. The proposed system need to perform a Natural language processing to extract Hotels features based on previous guests opinion. We have performed following four steps to process a Natural language text review as under:

- Lexical analysis
- Syntax Analysis
- Semantic Analysis
- Feature Extraction

*3.1.1. Lexical Analysis*

In Lexical analysis, the streams of characters are taken as input and streams of tokens are generated as an output.

a) Tokenizing

The hotel reviews are available in the form of paragraph which contains number of sentences or strings. These strings are tokenized into tokens or lexicons. These tokens are usually words, but can also be numbers or symbols. Usually all whitespace characters are removed from the sentences and alphabets or numbers are considered as a single token. These tokens are further gone through POS tagger to get different parts of speech called as morphemes. For example a verb "feels" is stored as "feel+s" and a noun "vegetables" is stored as "vegetable+s". Afterwards morphemes are lexically analyzed by a parse tree.

*3.1.2. Syntax Analysis*

In this step of Analysis, all the sentences and the phrases of the paragraph of reviews are authenticated in consultation with the defined grammatical rules in the English language. This paper uses Google Spell Check to correct the grammatical errors and typos in the crawled reviews. The misspelled words in review text are corrected by using a statistical spell checker (http://norvig.com/spell-correct.html) and remove the duplicate, unnecessary punctuation marks in sentences e.g., '!', ? etc. We have used the publicly recognized POS tagger to remove some noisy information contained in the review text such as syntactical errors and mistakes. The principal parts of all sentences are also identified in this phase. i.e. Object part, subject part, verb part, etc. Parse tree and typed dependencies are also generated in this phase. Syntactic dependency parser (http://nlp.stanford.edu/software/lex-parser.shtml.) can also return the syntactic dependency relations between words in a sentence.

a. Word Stemming:

There are many words which are derived from actual words called as derived words. Stemming is used to reduce derived words into their root forms. Lancaster Stemmer has been used in our work. Python NLTK provides WordNet Lemmatizer that uses the WordNet Database to lookup lemmas of words. The part of speech is first detected before getting the actual root word. In this process the part of speech of a word is first determined and different normalization rules are applied for each part of speech.

b. Extraneous word Removal

Reviews usually contain words which do not have significant meaning in extracting features for any product or item. such as 'the, 'a', 'also','about','an', 'at', 'to' etc. These words are removed. There is not any globally approved library for list of stop words present in the english language. To overcome this issue, we have developed library for such words in java of our own.

c. Shorten Exaggerated word
d. Words which have same letter repeating more than two times in a single word and not present in the lexicon are shorten to the meaningful word with the repeating letter occurring only once like exaggerated word "NOOOOOO" is reduced to "NO".
e. Part of Speech Tagging

The words with similar grammatical properties are classified through Part of speech tagging system as shown in Table 1 and Table 2. Each word in the review is separated and tagged according to part of speech it belongs to. The words are tagged as Singular Nouns, Plural Nouns, Verbs, Adjectives, and Adverbs etc. The NLP package returns tags like NN stands for singular common nouns and NP stands for singular proper nouns etc.

**Table 1.** Lexical analysis of the review

| | |
|---|---|
| **Review Strings** | ***Rooms of the hotel are big. Food is delicious. Hotel location is best. There is no internet.*** |
| **POS Tagging** | `The/DT Room/NN of/IN the/DT hotel/NN is/VBZ big/NN ./. Food/NN is/VBZ delicious/JJ ./. Hotel/NN location/NN is/VBZ best/JJ./. There/EX is/VBZ no/DT internet/NN` |

**Table 2**. Parse tree of a review sentence

```
(ROOT
    (S
      (NP
        (NP (DT the) (NN Room))
        (PP (IN of)
          (NP (DT the) (NN hotel))))
      (VP (VBZ is)
        (NP (NN big)))
      (. .)))
```

*3.1.3. Semantic Analysis*

In the semantic analysis all the tagged words among the sentences of the review are extracted in some sort of tabular form. It is decided in this particular stage that what actions are performed by the particular subject and a number of attributes related to every object are also identified. Output of the Semantic analyzer as in Table 3 contains a semantic table which is generated by the input review text on the basis of the parse tree generated in the previous phase.

**Table 3**. Semantic analysis

| No. | Tagged words | value |
|---|---|---|
| 1 | room | big |
| 2 | food | delicious |
| 3 | location | best |
| 4 | Internet | Not available |

If there is a noun in the sentence then we take it as a feature and store current sentence under this feature. The extracted information in the semantic table is used to express the features of the hotel. To find the polarity value for all the features from the user reviews using opinion mining approaches, a hotel feature matrix is obtained as in Table 4.

**Table 4**. Hotel x Feature Matrix after NL processing

| No. | Hotel-id | Review-id | Location | Price | room | Food | Staff |
|---|---|---|---|---|---|---|---|
| 1 | Hotel-1 | R-1 | 2 | 2 | 2 | 1 | 2 |
| 2 | Hotel-1 | R-2 | 1 | 1 | 1 | 2 | 1 |
| 3 | Hotel-1 | R-3 | 0 | 3 | 2 | 1 | 3 |
| 4 | Hotel-2 | R-1 | 2 | 0 | 2 | 1 | 2 |
| 5 | Hotel-2 | R-2 | 1 | 1 | 3 | 2 | 1 |
| 6 | Hotel-2 | R-3 | 2 | 2 | 2 | 1 | 2 |
| 7 | Hotel-3 | R-1 | 1 | 1 | 2 | 2 | 2 |
| 8 | Hotel-3 | R-2 | 0 | 1 | 1 | 2 | 1 |

*3.2. Polarity detection*

We identify the polarity of each review in the collection reviews using the NLTK library and calculate aggregated polarity score for each feature based on each review for every hotel from selected websites. We have started conducting feature based opinion mining of every review, where opinion indicates positive, neutral, or negative sentiment that a reviewer expressed on a feature based on opinion words. As there are multiple opinion words (great, nice, awesome) that are related with each feature (location, room, food) in a review. We assess every opinion word's sentiment strength which is also called polarity value. In this analysis, the values of review features and their associated opinions in terms of polarity are derived and are shown in Table 5.

**Table 5.** Polarity matrix

| No. | Review-id | Worst | Great | Shame | Awesome | Nice | Label |
|---|---|---|---|---|---|---|---|
| 1 | R-1 | 2 | 1 | 2 | 1 | 1 | Negative |
| 2 | R-2 | 1 | 1 | 1 | 2 | 1 | Positive |
| 3 | R-3 | 0 | 3 | 2 | 1 | 3 | positive |
| 4 | R-1 | 2 | 0 | 2 | 1 | 2 | negative |
| 5 | R-2 | 1 | 1 | 3 | 2 | 1 | Neutral |

*3.2.1. Generate TF-IDF*

After preprocess of textual reviews and removing all repetitive entries and unnecessary stop words, Tf-Idf (Term frequency—Inverse document frequency) needs to be generated for each review. We composite weights of each item in the review using Term Frequency-Inverse Document Frequency (TF-IDF) technique to determine which terms might be the most representative and occurring frequently in the collection of documents as well as which words are less representative and rarely occurring. TF-IDF is computed for each term word occurring in the collection of reviews. TF(t) is define as in Equation

$$TF\ (t) = \frac{Number\ of\ times\ term\ t\ appears\ in\ a\ document}{Total\ number\ of\ terms\ in\ the\ document} \quad (1)$$

While IDF for a term (t) is given as follows in equation

$$IDF\ (t) = Log \frac{Total\ number\ of\ documents}{Number\ of\ documents\ consisting\ term\ t} \quad (2)$$

Now we have to weigh down the frequent terms and find out the rare ones, by computing TF-IDF weight. The Tf-IDF weight is the product of TF (t) and IDF (t).

$$TF - IDF\ weight = TF\ (t) * IDF\ (t) \quad (3)$$

Sentiwordnet is a dictionary that tells, rather than the meaning, the sentiment polarity of a review. For detecting the polarity and subjectivity of different hotel reviews, and to get the polarity and subjectivity, we have used SentiWordNet, a publicly available analyzer of the English language that contains opinions extracted from a wordnet database. We separate our collection of reviews to extract words (hotel features) and assigned all representative occurring under the appropriate hotel features as explained in previous steps, find positive (pos), negative (neg) and neutral (neu) terms to calculate the sentiment score. SentiWordNet is included with Python's NLTK package and provide WordNet synsets with sentiment polarity. WordNet gives different types of semantic associations between words, which are used to calculate sentiment polarities. In simple words, sentiment analysis is the process of quantifying something which is qualitative in nature such as textual reviews. The sentiment score of a term (pos or neg) is multiply by TF-IDF weight to calculate overall sentiment score (polarity) of terms in the document. The following equation.

$$overall\ sentiment\ (polarity) = sentimentscore * TF - IDF\ weight \quad (4)$$

The overall sentiment polarity score (negative or positive) explains that how many features are positively or negatively important in the hotel review. As there are multiple opinion words that are related with each feature in a review, a weighted average value is calculated which acts as the weight to represent the overall positive or negative polarization of the review. If the polarity score of a feature in the reviews of a hotel is greater than zero then the feature is the positively polarized and if it is less than zero then it is negatively polarized and if it is equal to zero then it represents the neutrality. We have calculated the polarities of all the reviews of the hotels taken from different data sources.

The polarity of the reviews *Pr* of a hotel from the selected website can be calculated by taking the difference of the aggregated polarity score of positive reviews *posr* and the aggregated polarity score of negative reviews *negr* of that particular hotel $h_n$ from the particular selected website $w_o$.

$$Polarity_{rm}\ (P) = sgn \left[ \left| \sum_{m=1}^{n} (posr_m) \right| - \left| \sum_{m=1}^{n} (negr_m) \right| \right] \quad (5)$$

Where $posr_m \wedge negr_m \in h_n$ and $h_n \in w_o$

Then, we will take aggregated polarity score of textual reviews of each hotel from each selected website. Aggregation is the process of combining things. That is, putting those things together so that we can refer to them collectively.

$$aggregated\ Polarity_{hn}(A) = \sum_{m=1}^{n} P_{rm} \tag{6}$$

The weighted average of the aggregated polarity by total reviews of the respective hotel from the selected website and the weight score of ranks and votes will be calculated as in below equation.

$$weigted\ average\ Polarity_{hn}(B) = \frac{A_{hn}}{T} + (Aggregated\ ranks_{wo}) + (Aggregated\ votes_{wo}) \tag{7}$$

Here $T$ is the total number of reviews of hotel $h_n$ from hotel website $w_o$.

$$aggregated\ weighted\ average\ Polarity_{hn}(C) = \sum_{O=1}^{N} (B_{wo}) + (Likes_{hn}) \tag{8}$$

$$average\ aggregated\ weighted\ average\ Polarity_{hn}(D) = \frac{C_{hn}}{N} \tag{9}$$

$N$ is the total number of selected hotel websites containing large number of hotel reviews.

$$Final\ score_{hn}(F) = \frac{r_{hn} - min(r_{hn})}{max(r_{hn}) - min(r_{hn})} * 10 \tag{10}$$

Here $F$ is the final value of the normalized average aggregated score $r_{hn}$ of the hotel ($h_n$) from number $N$ of selected websites

*3.3. Type classification and recommendation*

The classification is done by calculating the final score. The reviews words are matched with the dictionary words and if it is a positive word then score will be +1, if negative word then score will be -1, otherwise 0. The final recommendation is achieved by using the fuzzy logic approach. The fuzzy sets theory provides a framework for the representation of the uncertainty of many aspects of human knowledge. For a given element, fuzzy sets theory presents the degrees of membership to a set. For example; if we have the set of solo guests then we can consider that a person who likes to do gym or take massage in hotel belongs to such a set with a degree of 1 or a person who likes to have a ghazal night or cinema in the hotel must be a couple guest and belongs to a set in some other degree. The purpose of our recommender is to provide hotel recommendations based on some expert criteria using the fuzzy set. The first step of the recommendation process consists of representation of knowledge about how the hotels are selected. This knowledge is expressed using fuzzy sets.

In the first step, the fuzzy sets are defined based on the expert knowledge. The expert explains the characteristics of the hotels and the characteristics of the customers in terms of fuzzy sets. In the second step consists of doing recommendations using a previously build hotel-feature-rating matrix which is usable by a recommender system based on a collaborative filtering technique.

*3.3.1. Fuzzy set*

To represent the degree of membership of a certain hotel to a certain class, fuzzy sets theory is used. The final recommendation is achieved by using the fuzzy logic approach based on the fuzzy rules by calculating the final score to provide the class of the hotel based on guest type (solo, couple etc.) as shown in Figure 4. Fuzzy rules are defined as follows.

Rule 1: if F > 8 then Hotel Type "R (Recommended)".
Rule 2 else if F > 6 and F ≤ 8 then Hotel Type: "BR (Best Recommended)".
Rule 3 else if F > 4 and F ≤ 6 then Hotel Type: "AR (Average Recommended)".
Rule 4 else if F > 2 and F ≤ 4 then Hotel Type: "LR (Least Recommended)".
Rule 6 else Hotel Type: "NR (Not recommended)".

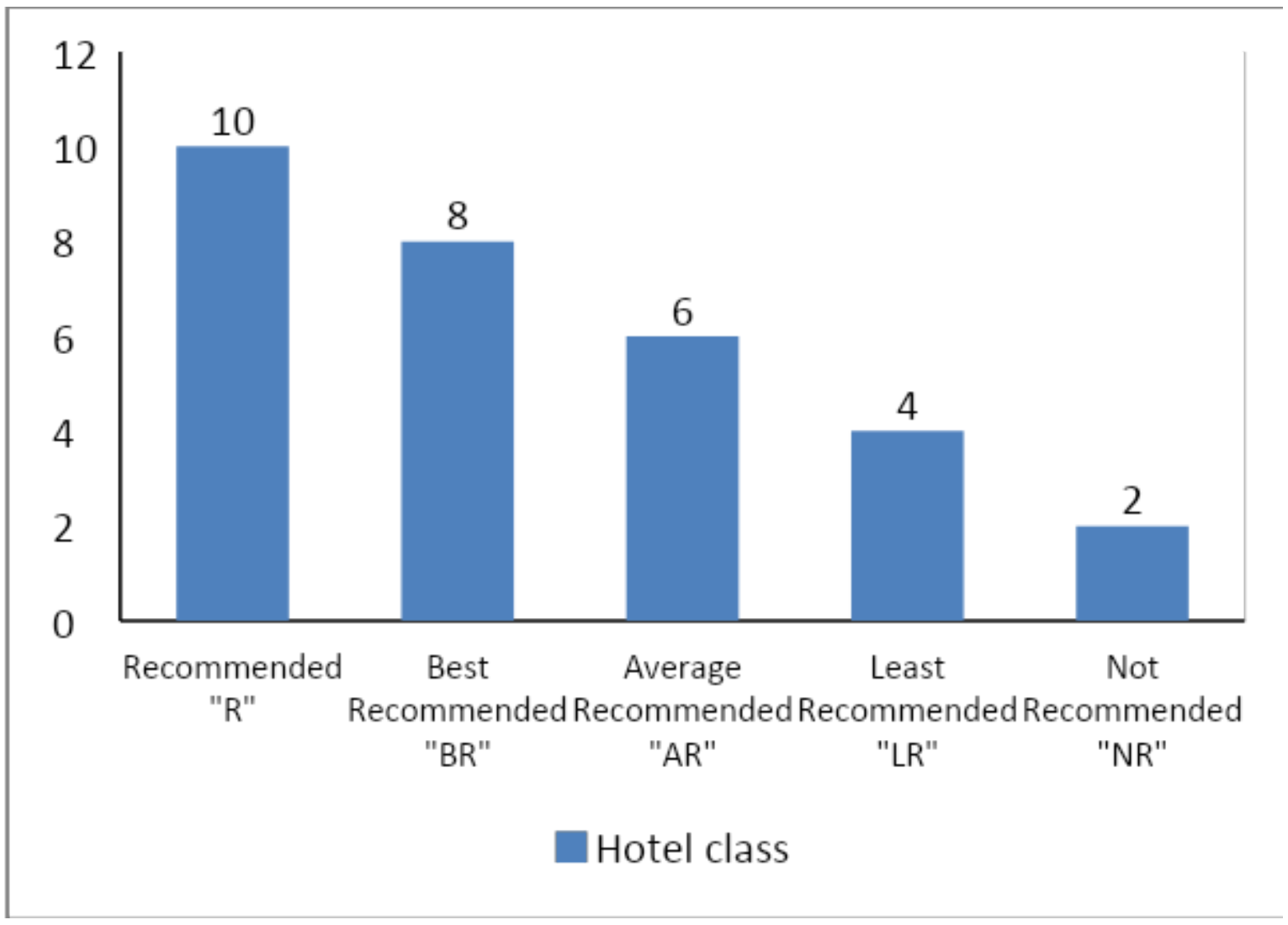

**Figure 4**. Hotel recommendation class

## 4. Implementations setup

### *4.1. NoSQL Storage*

A recommendation architecture needs to store diverse sorts of data. Cassandra has been used to store our data. Review pages which match the Query keywords are downloaded and are stored in the NoSQL database in Hadoop. In the first place, we have to manage a large number of data of different users such as users' ratings as well as reviews. The dataset used in this study is crawled in from the hotel website of TripAdvisor and Expedia. The information of hotel title and other identifying information are setup as in the comma separated value format. Data set is converted from CSV to the JSON format to make it easier to read. The users' textual reviews and ratings assigned by existing users recorded as i.e. rating score, likes or star ranks, are stored in Cassandra. The ranks score can vary between the different scales of 1 to 5 or 1 to 10. Normalized ranks are calculated in this paper. The use of Cassandra in the proposed system not only improves the response time of the web application but these improvements are noticeable in terms of reduced recommendation response time. The dataset used in this study is taken from the hotel websites. The metadata includes hotel Records and customer profiles which are used in experiments for numerical simulation tests in Cassandra. The data collection process is shown in Figure 5.

In our proposed application, the working process of the proposed recommender can be described by the following steps.

- Start the process.
- The active user provides its choice as an input as per given search criteria based on the guest type such as solo, couple, business etc.
- Then the system checks for the previous users' data (Ratings, ranks, reviews) matching the query from the web in the system database.
- The system filters the query data by matching query in the external web sources available.

- If match with query, then collect metadata.
- Save the metadata in the NoSQL database.
- If do not match. Then discard it.
- Repeat until all matched metadata is found.
- End

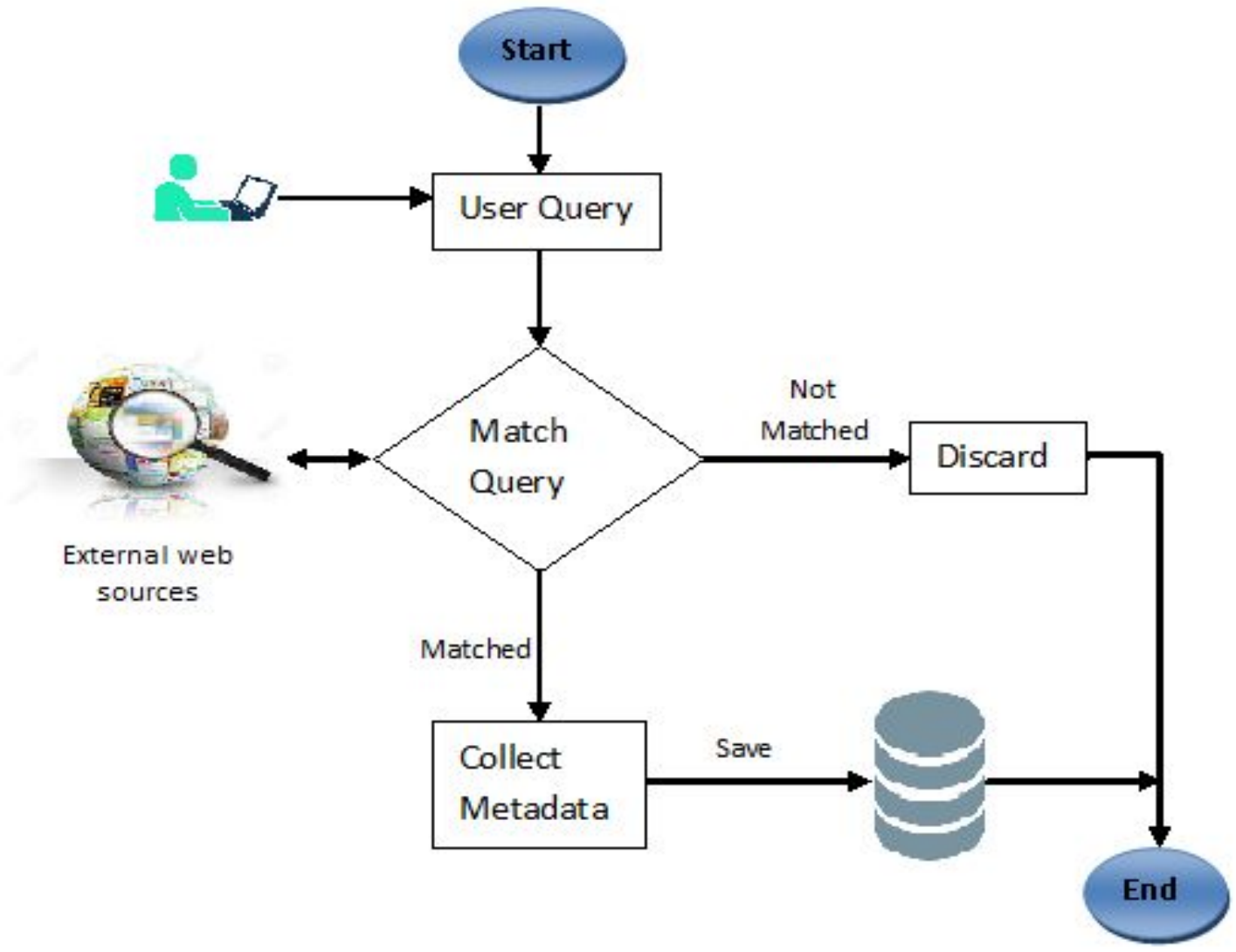

**Figure 5**. Data collection

*4.2. System Overview*

The architecture of the proposed Recommender system application consists of three main components such as external resources, Front end and backend as shown in Figure 6. The dataset containing hotel reviews and ranks, is taken from the external Hotel websites of TripAdvisor and Expedia. Reviews are divided into different data sets to check the performance and working of our methodology in Hadoop using Cassandra After the completion of this process, the complete data is stored into system databases using web crawler written in java in the developed methodology.

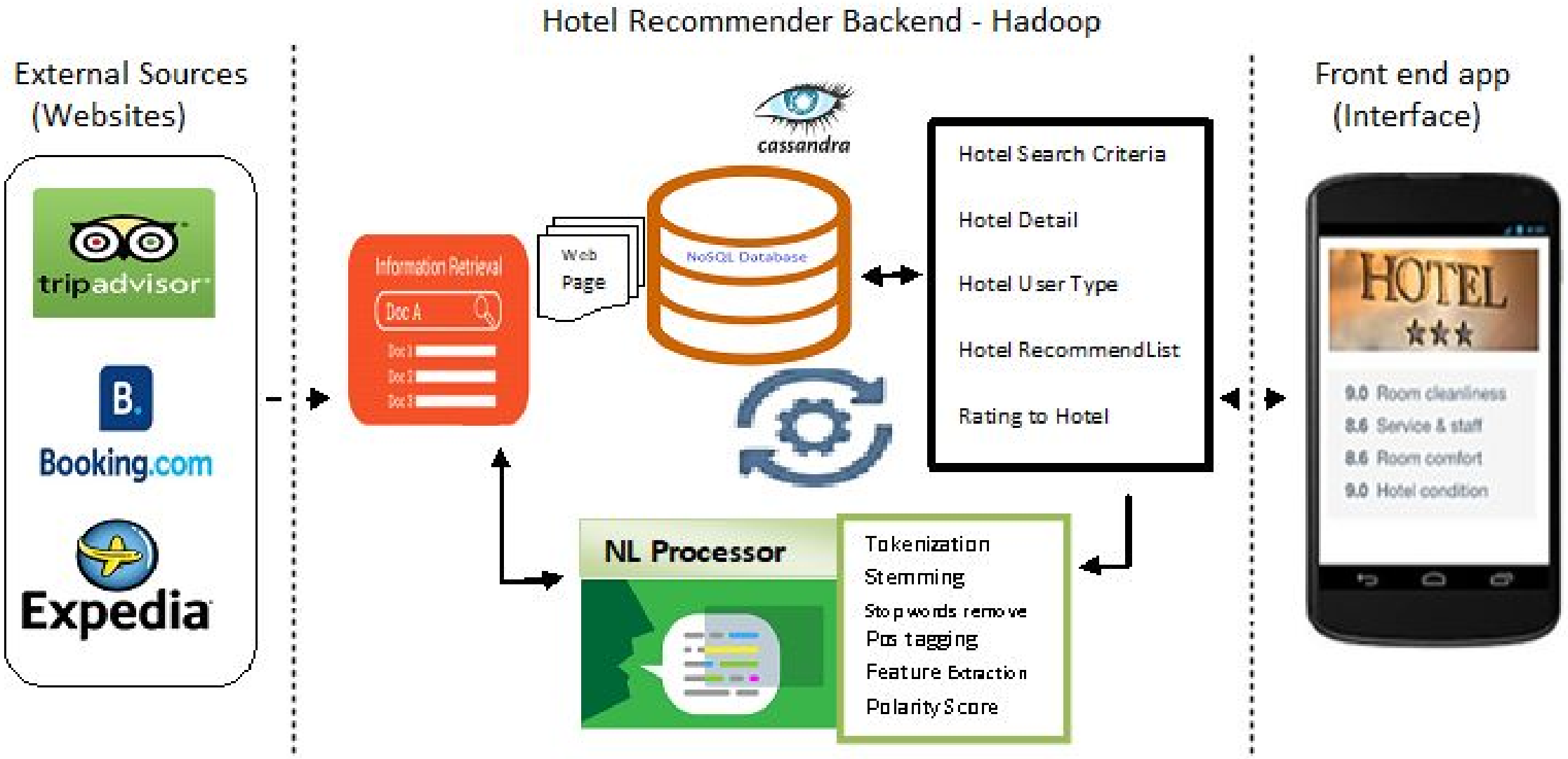

**Figure 6.** Proposed hotel Recommender Architecture

In order to get best recommendations regarding the desired hotels, the proposed hotel recommender system is designed in Python and Java code with templates of Smarty. The application is supportive to any web browser and the user can access it online from any platform, it uses development environment based on reliable open source tools. The computing environment is also discussed in Table 6. In the hotel recommender Application, every search parameter has been interfaced so that users can query according to their choice against each search criteria. The hotel recommender is designed to provide hotel recommendations using the reviews polarity scores and ratings calculated using the reviews users data stored in Cassandra which matched the active user query.

### *4.3. Computing Resources*

The proposed system has been tested in the development environment which is based on reliable open source tools such as Cassandra, JSON, PHP, WordNet and Java. The computing environment details are given in Table 6.

**Table 6.** Recommender System Computing Environment

| **System Specification** | |
|---|---|
| **Hardware** | **4 Core(TM) i7-3770, 3.40GHz,500GB Disk, 8GB RAM** |
| **Software** | **CentOS 7.0, Hadoop 2.0, Apache Cassandra 3.11.1, JDK 1.8/Eclipse, and PHP 7.1** |
| **Plug-In** | **PHP-Cassandra extension** |

### *4.4. Web Service*

The hotel recommender system can be accessed via a web page. First of all, a user can connect through specified URIs and HTTP connections. After that when the user requires any data from the web page, a request is submitted through HTTP method along with the necessary parameters required in a particular method and the requested data is displayed through the web page. The response message of HTTP is converted into the JSON format to keep the uniform data format. The description of used functions along with their query parameters in URIs is shown in Table 7.

**Table 7.** Description of Methods Applied

| **Name** | **Explanation** | **Other Query Parameters** | **HTTP Request Method** |
|---|---|---|---|
| **Search** | **This method scrutinize data using search criteria (Name/Rating/City/Region to get the list of available hotels data** | ***Searchtitle*** | **GET** |
| **getratings** | **Need previously assigned users ratings before returning the list of recommendations showing percentage.** | ***Ratingsinnum*** | **GET** |
| **dispratings** | **Show the specified hotels with rating** | ***ratingstars*** | **GET** |
| **hotel Recommendation** | **provides a list of recommended hotels** | ***Id/name*** | **GET** |
| **hoteldetail** | **Give hotel details with name id and region** | ***Id*** | **GET** |

The recommender system is accessed using a URL (http://HotelRecommenderSystem/User_request/?method) is used to convert data contents in

JSON to deliver the particular information about the hotel. JSON is used to make it easier to read the data. The most significant point of interest is that the web services over the recommender are protected by SSL during the data transmissions.

*4.5. Experiment and Results*

We have used two reliable data repositories (TripAdvisor, Expedia) containing significant number of ranks, ratings and reviews to represent heterogeneity of data i.e. textual reviews and numerical ratings and ranks. These data sorces contain data for 4000 of the most popular hotels and collection of hotel reviews and ratings which is useful in our experiments. After data preprocessing, tf-idf generation and polarity deection using SentiWordNet, we have computed the polarity scores for textual reviews obtained from each selected website. We have illustrated the data obtained from the selected hotels and the corresponding data source and its processing in the proposed methodology in Table 8.

**Table 8.** Polarity Scores, Rank scores and voting scores

| Data Source / Selected Hotel | TripAdvisor (D1) | | | Expedia(D2) | | |
|---|---|---|---|---|---|---|
| | **Polarity Score** | **Normalized rank Score** | **Voting Score** | **Polarity Score** | **Normalized rank Score** | **Voting Score** |
| **SpringHill Suites Denver Downtown (H1)** | 31 | 3.9 | 209301 | 23 | 5 | 268231 |
| **Amsterdam Court Hotel (H2)** | -5 | 3.4 | 38821 | 7 | 4 | 63420 |
| **Mandarin Oriental New York (H3)** | 17 | 3.2 | 111620 | 24 | 5 | 127023 |
| **Millennium Hilton (H4)** | -4 | 2.8 | 17023 | -3 | 3.5 | 35622 |
| **Belnord Hotel (H5)** | 16 | 3.5 | 29441 | 22 | 5 | 41323 |

In table 8, the selected external data source used in our work such as TripAdvisor is represented by “D1” and Expdia is represented as “D2”. Similarly the corresponding selected hotels are represented as Mandarin Oriental New York “H1”, Amsterdam Court Hotel “H2”, Hotel Metro “H3”, Millennium Hilton “H4” and Belnord Hotel “H5”. We have calculated polarity scores of the textual reviews for each hotel from the selected websites. Normalized rank score (scale 1-5) and the voting score are also calculated. Voting score represents the number of votes given by the customer to each hotel. Hotel names and Data Sources with their corresponding IDs are shown in the table.

The Normalized rank scores of a selected hotel from two selected data Sources such as TripAdvisor “D1” and Expedia “D2” are plotted in the Figure 7 and it has been noted that the Expedia has high ranks comparatively in comparison with TripAdvisor.

Polarity scores of textual reviews of hotels taken from different selected data sources are calculated and aggregated in Table 9. Weighted average polarity is achieved by adding normalized ranks and votes in the average polarity (calculated by taking average of aggregated polarities).

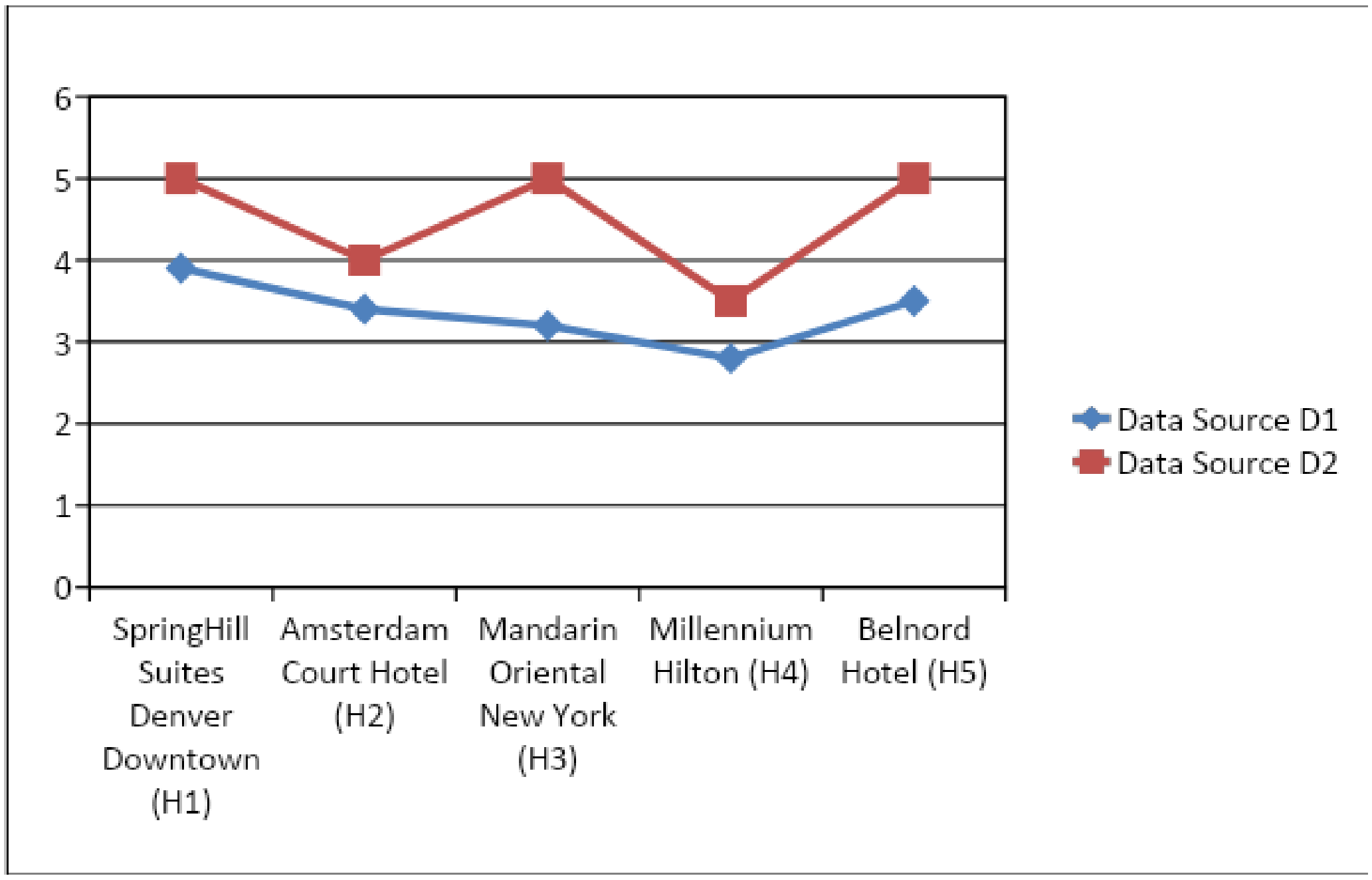

**Figure 7.** Difference in ranks scores

The power of social media websites such as twitter, YouTube and facebook is also creating a shift in the way travelers seek out suggestions and tips before making any booking decision for certain hotel. Videos contain hotels pictures as well as present hotel services which may also affect the behavior of customers before selection of hotel and also have impact on the hotel rating. That is why we have used YouTube video views in our work to present the heterogeneous approach. We added number of views in weighted average polarity to calculate the aggregated weighted average polarity for quality recommendations. The Final scores along with hotel classes are shown in Table 10.

**Table 9.** Weighted average polarity computation

| | Reviews | | Aggregated Polarity | | Average Polarity | | Weighted Average Polarity | |
|---|---|---|---|---|---|---|---|---|
| Hotel | D1 | D2 | D1 | D2 | D1 | D2 | D1 | D2 |
| H1 | 1309 | 1519 | 31 | 23 | 0.023 | 0.015 | 209304.92 | 268236.01 |
| H2 | 396 | 456 | -5 | 7 | 0.012 | 0.015 | 38824.41 | 63424.01 |
| H3 | 1189 | 998 | 17 | 24 | 0.014 | 0.024 | 111623.21 | 127028.02 |
| H4 | 537 | 337 | -4 | -3 | 0.007 | 0.008 | 17025.80 | 35625.50 |
| H5 | 971 | 1117 | 16 | 22 | 0.016 | 0.019 | 29460.50 | 41328.01 |

These heterogeneous data source such as ranks, votes, textual reviews and views are computed using the proposed approach and final rank scores are obtained as shown in Figure 8. The hotels are classified based on the final ranking score. The same is explained in Table 10 and hotel class is identified based on the fuzzy set used in our work.

**Table 10.** Computation of final score

| Hotel ID | YouTube Views | Aggregated Weighted Average Polarity | Final Score | Hotel Class |
|---|---|---|---|---|
| H1 | 331025 | 569795.465 | 8.93234 | R |
| H2 | 89023 | 140147.21 | 3.92215 | LR |
| H3 | 284230 | 403555.615 | 7.63217 | BR |
| H4 | 56056 | 82381.65 | 2.08245 | LR |
| H5 | 78124 | 113518.255 | 3.12871 | LR |

The Final rank score “F” of H1 hotel “SpringHill Suites Denver Downtown” is greater than 8 that is why it is placed in class “R”. The H3 hotel “Mandarin Oriental New York” lies in “BR”class because “F” score is greater than 6 and is less than 8. Whereas, the H2 hotel “Amsterdam Court Hotel”, H4 hotel “Millennium Hilton “ and H5 hotel “Belnord Hotel “ score is greater than 2 and less than 4 so it lies in class “LR”.

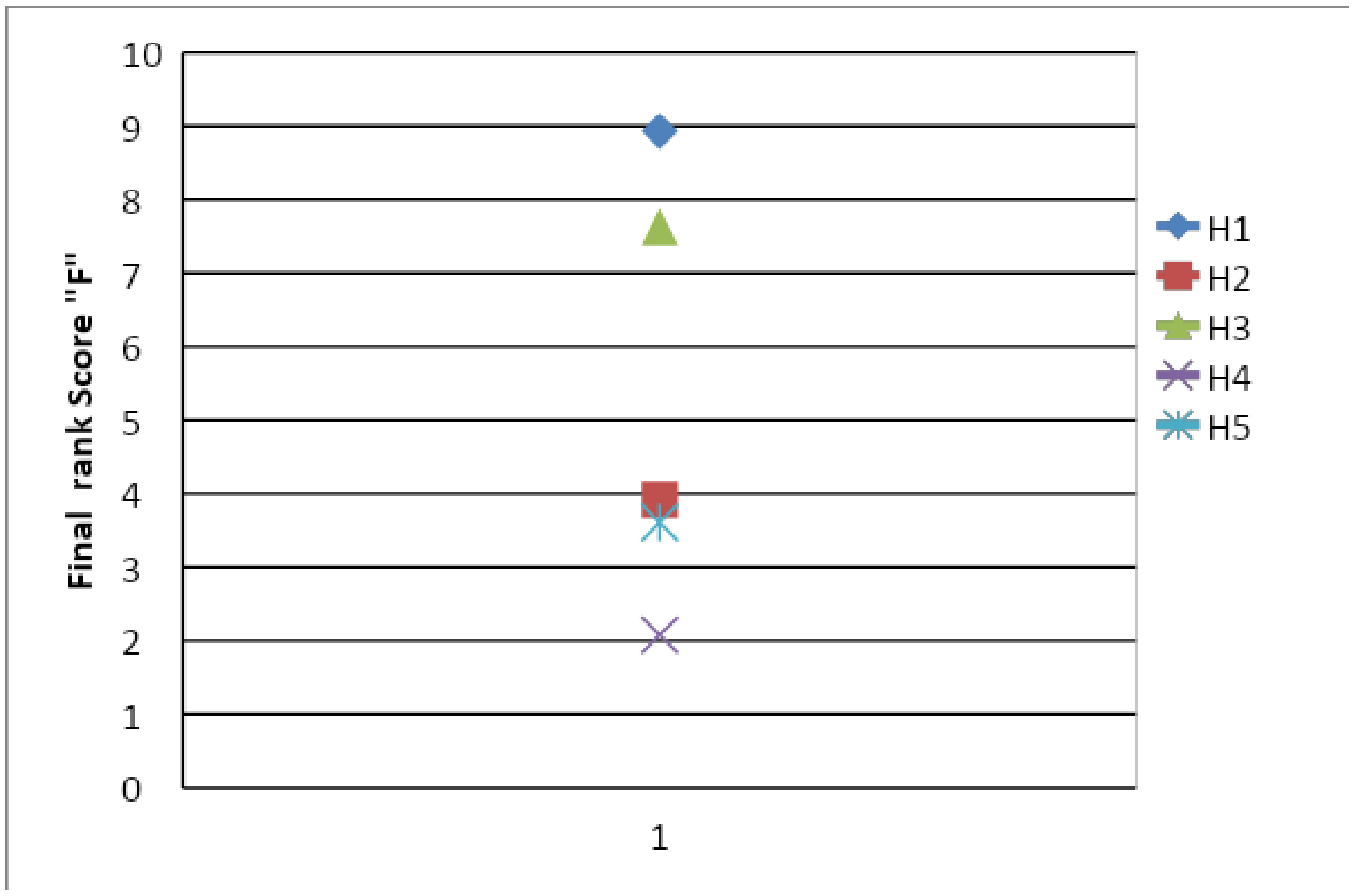

**Figure 8.** Proposed hotel recommender final ranking

Figure 8 represents the class of each hotel taken from selected data sources against the final rank score computed based on the guest type such as solo, family, couple etc.

Figure 9 represents the each recommended criteria Score against the selected data source and the proposed system using Heterogeneous data.

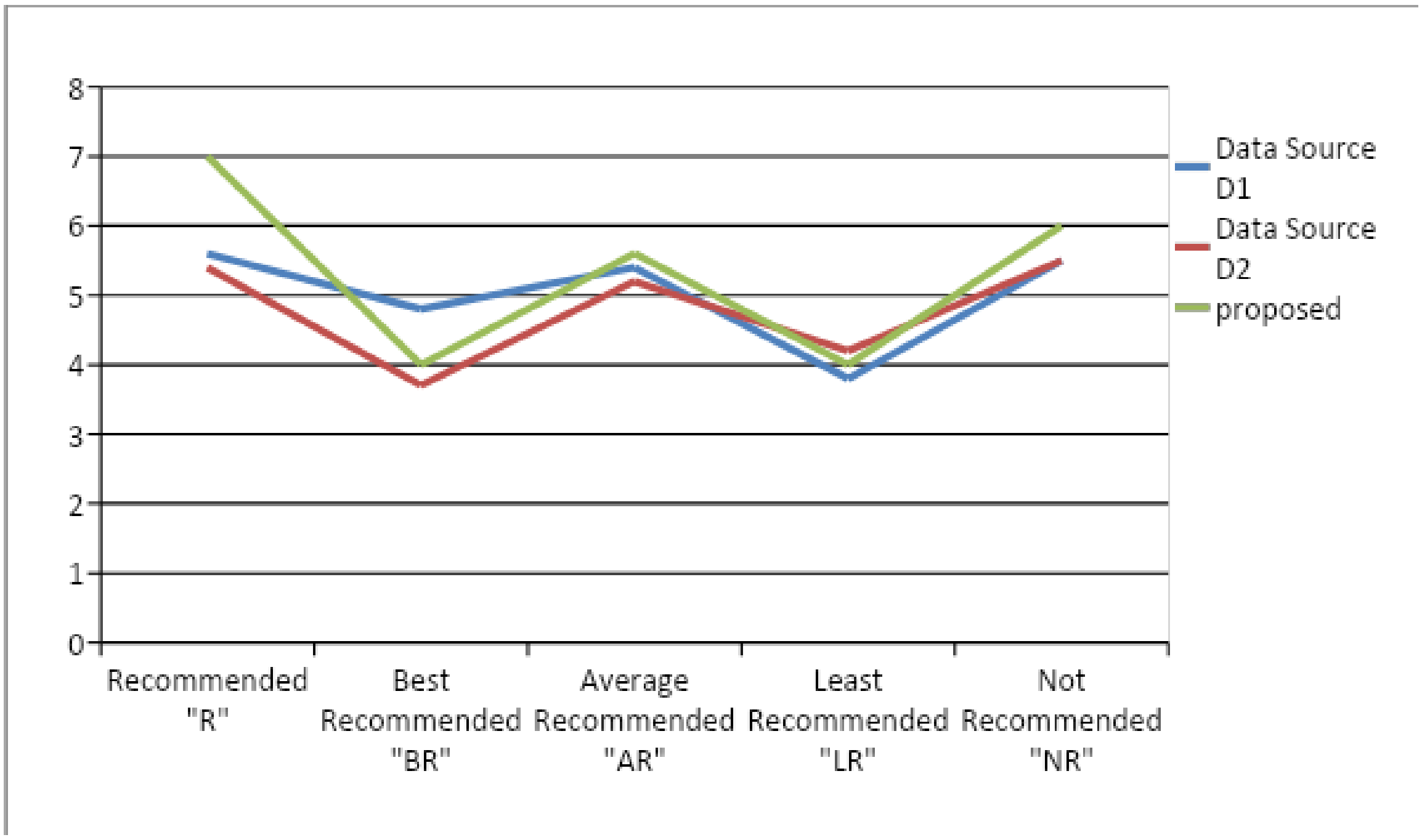

**Figure 9.** Comparison of hotel scores

This developed hotel recommender service gives the users the opportunity to explore the collection of hotels by applying several filters such as hotel name, location and user Ratings. The system is able to filter the hotels on the basis of system define criteria depending upon the guest type. The list of recommended hotels generated by the proposed system is displayed in Figure 10.

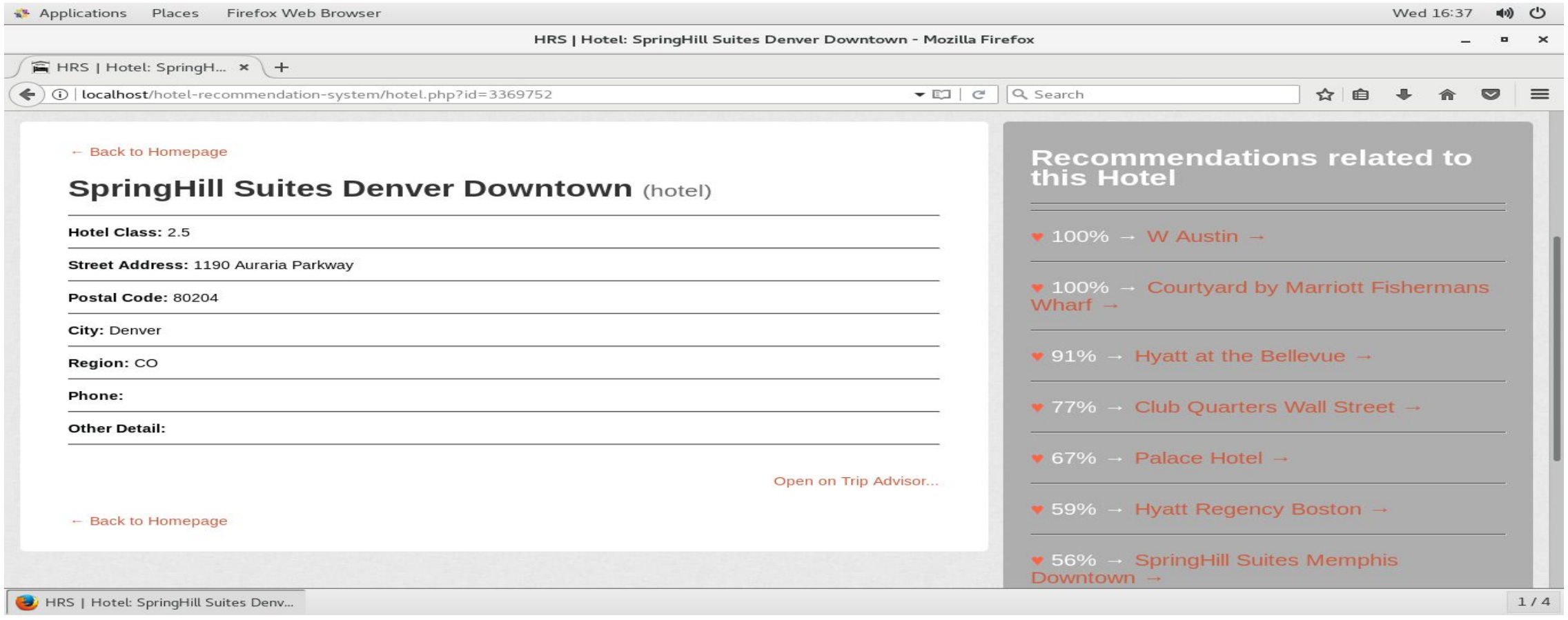

**Figure 10.** List of recommendations

## 5. *System Evaluation*

Once the recommender has been proposed and developed, it is essential to demonstrate the accuracy of our proposition. The system should demonstrate that the recommendations given by the proposed system are acknowledged by the targeted users. By considering the end goal to realize whether the proposed Recommendation application is compelling and effective to resolve scalability by the framework, the three exemplary accuracy measurements tools are utilized as assessment measurements. The first measure is called as the Precision rate which is represented as below

$$Precision = \frac{Z}{X+Z} \times 100\% \quad (11)$$

As indicated by user inputs in the framework, we assume that Z is the quantity of recommended hotels which the user is interested in and X represents the number of hotels that the user is not interested in, 'X+Z' represents the total number of hotels that the system suggests/recommends to the user. To calculate the ratio on the number of the recommended hotels that are liked by the user, over the aggregate sum of the hotels in the system in which the user is possibly interested. Again keeping the above assumptions the Recall rate will be represented by the below formula.

$$Recall = \frac{Z}{Y+Z} \times 100\% \quad (12)$$

In the equation (16) 'Y' represent the number of hotels targeted, however not being recommended to the user, and Y+Z indicates the list of all hotels in which the user may possibly be interested in. However, the Precision measure and the Recall measure are contrary to each other while measuring the accuracy. Perfect Recall rate of "1" means comparatively low Precision rate. If the recommender suggests only one hotel that user may like, the precision will be a perfect score" 1" but the recall will be closed to "0" in the large scale hotel pool. On the contrary, a perfect recall of "1" will happen, though the precision rate will be very low, may users receive list of all interesting hotels in the recommender that they absolutely like. The third accuracy measure used here is called as F-measure. It is represented by the following formula.

$$F - measure = 2 \times \frac{Precision \times Recall}{Precision + Recall} \quad (13)$$

Above two metrics represented by equation (11) and (12) are used to calculate F-measure represented by equation (13), F- measure is a weighted combination of the Precision and recall metrics. The traditional F-measure is the Harmonic Mean of Precision and Recall and here it is used in this paper to evaluate the accuracy and efficiency of the recommendations in the proposed Recommender application.

**Table 11.** Evaluation Metrics

| **Incremental data update using Euclidean similarity** | F-measure | Recall | Precision |
|---|---|---|---|
| **20%** | 0..966 | 0.954 | 0.978 |
| **40%** | 0.956 | 0.941 | 0.972 |
| **60%-** | 0.951 | 0.936 | 0.968 |
| **80%** | 0.938 | 0.921 | 0.956 |
| **100%** | 0.924 | 0.898 | 0.951 |
| **Avg. ratio** | 0.950 | 0.930 | 0.965 |

Higher F-Score represents better quality of recommendations. Alongside these standard measures, we will moreover utilize an uncommon assessment convention to evaluate the results of the incremental update procedure. We have divided the metadata into training (60%) and testing (40%) datasets. Depending on these data parts, we initially performed the calculations on the underlying 60% of the data of the complete metadata in the form of chunks of 20% of data for training of the developed system and computed Precision, Recall and the F-measure. Secondly, we incrementally included the metadata of the remaining of the 40% dataset into the system for testing the system. Repeated measurements of accuracy metrics Precision, Recall, and the F-measure with each increment are performed as shown in Table 11.

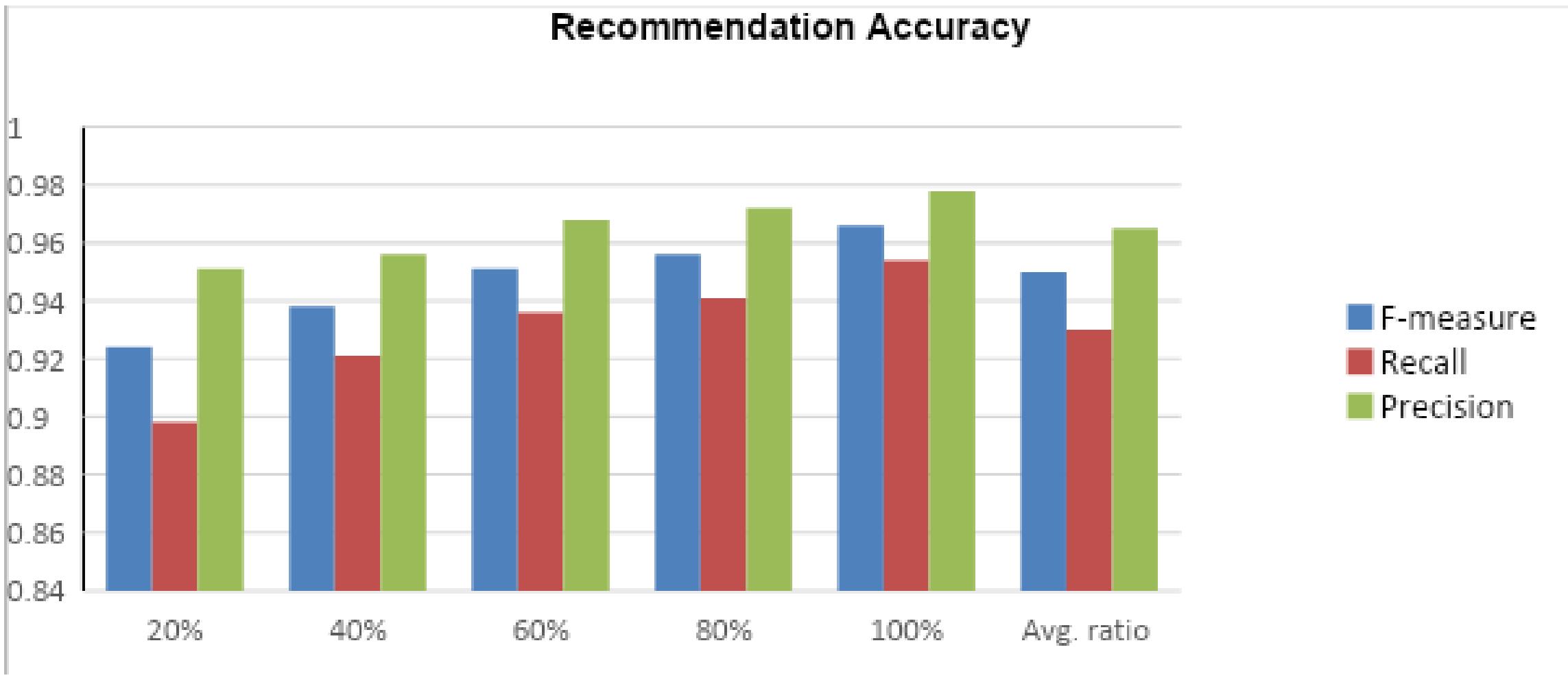

**Figure 11.** Performance metrics with Incremental chunks of Dataset

Figure 11 shows the F-measure values of recommendations under the proposed mechanisms. The proposed recommender system supports computations to evaluate how it would influence the recommendation accuracy. The Euclidean similarity in the system with NoSQL dataset and proposed opinion mining approach provides accurate recommendations, and its F-measure ratio value is 0.950. The total number of hotels in the used dataset is very huge, so the performance measures of Precision rate is bigger than the recall and F –measure rates which can clearly be seen. It shows that the Euclidean similarity in the recommender system shows accurate prediction to users.

We have gathered a group of 12 users of different age groups and provided them the chance to use our designed recommender system as well as other available recommenders such as yatra.com, trivago.com, and hotels.com to get recommendation for hotels of their own choices. These participants or users have been asked to give express their satisfaction level after performing searches over each of these recommenders. They have performed random searches and their satisfaction level is recorded in a table in three categories i-e less satisfied(LS), satisfied(S), highly satisfied(HS) against three parameters i-e time efficient(T), relevance of results with query(R), cost effective(C) as shown in Table 12.

Satisfaction level of the participants' shows that after using the proposed recommender, most of users has given best category to the proposed recommender system. Performance outcomes generated in terms of response time are recorded in Table 13 and shown in a Figure 12. The performance and response time (processing Time which includes fetch time, load time and search time) obtained after experiments to depict the performance of the system are graphically shown in Figure 12.

**Table 12.** Participants Opinions

| No. | Professio n | Age | Trivago.com | | | Yatra.com | | | Hotels.com | | | Proposed Recommender | | |
|---|---|---|---|---|---|---|---|---|---|---|---|---|---|---|
| | | | T | R | C | T | R | C | T | R | C | T | R | C |
| **User 1** | **Business** | **35** | **LS** | **LS** | **S** | **LS** | **LS** | **LS** | **S** | **LS** | **LS** | **S** | **HS** | **HS** |
| **User2** | **Student** | **21** | **S** | **S** | **LS** | **LS** | **LS** | **LS** | **LS** | **S** | **LS** | **HS** | **HS** | **LS** |
| **User3** | **Doctor** | **40** | **LS** | **LS** | **LS** | **S** | **LS** | **LS** | **LS** | **LS** | **LS** | **HS** | **S** | **HS** |
| **User4** | **Teacher** | **33** | **LS** | **LS** | **LS** | **LS** | **LS** | **S** | **S** | **LS** | **S** | **HS** | **S** | **HS** |
| **User5** | **Student** | **17** | **HS** | **S** | **S** | **LS** | **S** | **LS** | **LS** | **S** | **S** | **LS** | **HS** | **HS** |
| **User6** | **Business** | **52** | **LS** | **LS** | **LS** | **LS** | **LS** | **HS** | **S** | **LS** | **S** | **S** | **HS** | **S** |
| **User7** | **Business** | **26** | **S** | **LS** | **LS** | **LS** | **LS** | **LS** | **LS** | **S** | **LS** | **S** | **HS** | **HS** |
| **User8** | **Employee** | **37** | **LS** | **S** | **HS** | **LS** | **G** | **LS** | **LS** | **LS** | **LS** | **S** | **HS** | **HS** |
| **User9** | **Student** | **19** | **LS** | **S** | **LS** | **S** | **S** | **LS** | **S** | **LS** | **S** | **HS** | **LS** | **LS** |
| **User10** | **Teacher** | **44** | **LS** | **LS** | **S** | **LS** | **LS** | **S** | **S** | **S** | **S** | **HS** | **LS** | **S** |
| **User11** | **Teacher** | **50** | **LS** | **LS** | **LS** | **LS** | **LS** | **S** | **S** | **LS** | **HS** | **S** | **S** | **HS** |
| **User12** | **Doctor** | **30** | **S** | **LS** | **LS** | **LS** | **LS** | **LS** | **LS** | **S** | **LS** | **S** | **S** | **HS** |

When the user logs into the designed recommender application and query for best hotel recommendation providing a guest type according to their choice, the system performs computation using proposed approach to collect the required recommendations. The system has shown promising results by time reduction in the processing time.

The system also calculates the time stamps as loading time, searching time and execution time represented as milliseconds (ms). The sum of loading time and searching time together is called as execution time and is also presented graphically in Figure 12. Cassandra outperforms in combination with CF and takes very less time while searching.

**Table 13.** Response Time

| **No.** | **Load Time** | **Search Time** | **Execution Time(ms)** |
|---|---|---|---|
| 1 | 3.0994 | 0.0455 | 3.5544 |
| 2 | 1.6212 | 0.0500 | 1.6712 |
| 3 | 3.0994 | 0.0461 | 3.5606 |
| 4 | 2.8610 | 0.0738 | 2.9348 |
| 5 | 3.0994 | 0.0447 | 3.1441 |
| 6 | 1.0013 | 0.0488 | 1.0501 |
| 7 | 1.9073 | 0.4583 | 1.9531 |
| 8 | 2.8610 | 0.0469 | 2.9079 |
| 9 | 3.0994 | 0.0024 | 3.1018 |
| 10 | 2.8610 | 0.0459 | 2.9069 |
| 11 | 2.8610 | 0.0727 | 2.9337 |
| 12 | 2.1457 | 0.0461 | 2.1918 |

Total query execution time in milliseconds was 1265(1.265 seconds). The comparison of results in testing phase shows that the Cassandra NoSQL database is efficient and helps to reduce the total processing time. Results indicate that the system is performing efficiently which impact people opinions about using recommenders.

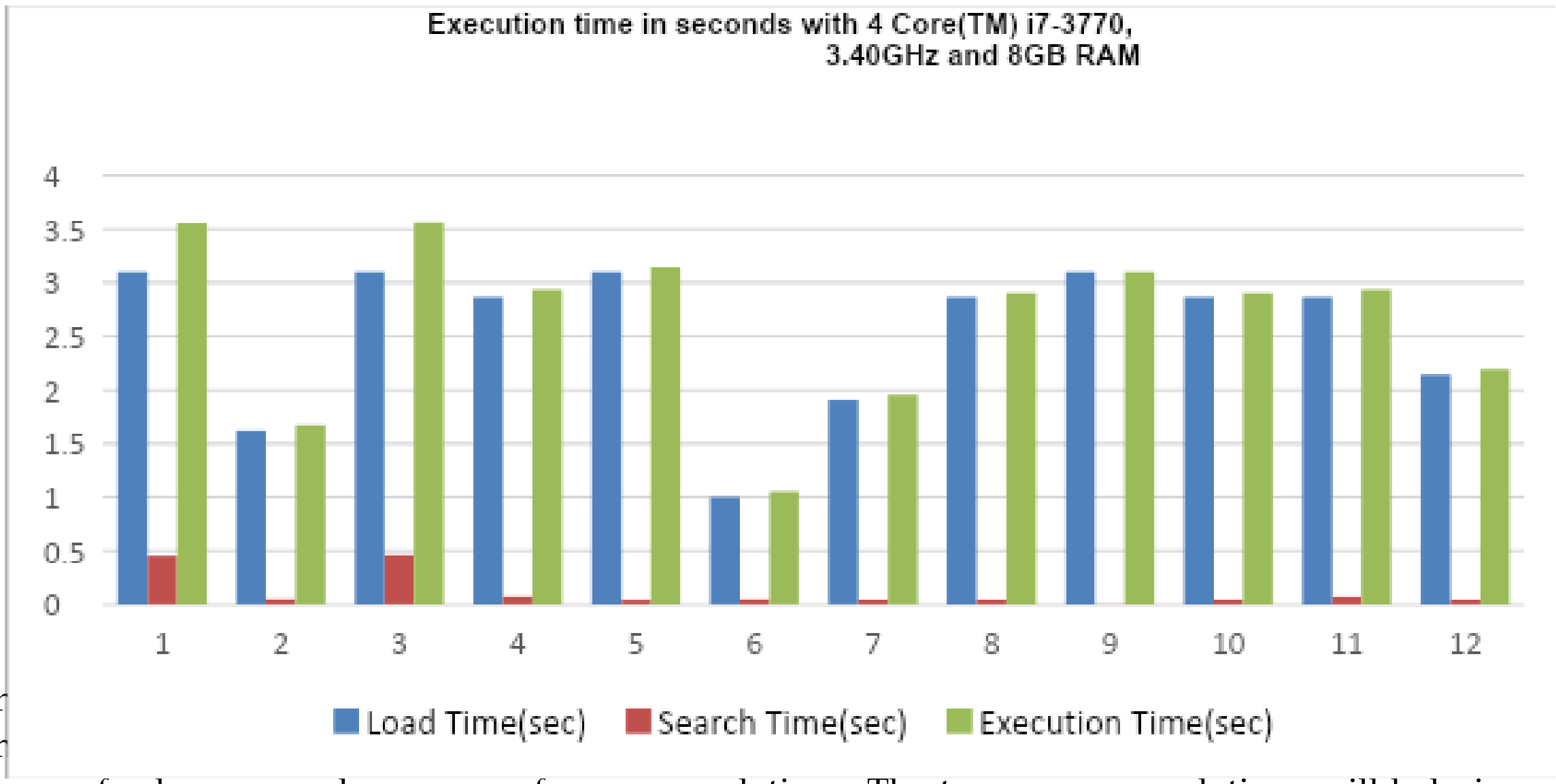

Th

im

terms of relevance and accuracy of recommendations. The true recommendations will help in improving the customer's satisfaction level and that will play role in success of e-commerce applications that will ultimately affect the business strategy and planning.

**5. Comparative Analysis**

The comparative analysis of the proposed Architecture is presented with existing research. The information filtering (IR) techniques already discussed in existing literature are compared. Similarity measures used along with different IR techniques and recommender algorithms are compared. For a statistical analysis, the F-measure is used along with precision and Recall measure in the proposed framework as an evaluation tool. Other Analysis tools used in the comparative studies are also analyzed. To evaluate prediction accuracy of the proposed method, we have performed the comparison of the performance analysis metrics such as recall, precision and F-measure of our hotel recommended system with the precision, recall and F-measure of traditional related recommenders included in this paper in Table 14.

**Table 14.** Comparison with related studies

| No. | Reference | Precision | Recall | F-measure |
|---|---|---|---|---|
| **1** | **Chang et al.** | **0.43** | **0.29** | **-** |
| **2** | **Liu et al.** | **0.56** | **0.60** | **0.59** |
| **3** | **Zhang et al.** | **0.25** | **0.34** | **0.35** |
| **4** | **Hsieh et al.** | **0.02** | **0.53** | **0.01** |
| **5** | **Proposed** | 0.965 | 0.930 | 0.950 |

We have also compared existing studies with our approach and find out promising improvement in terms of execution time of the proposed approach. The comparative analysis of the performance of the proposed hotel Recommender approach with the existing related approaches found in the literature is shown in Table 15 which exhibits outcomes in terms of time improvement.

**Table 15.** Comparison of execution time with related studies

| No. | Reference | Recommendation Time |
|---|---|---|
| 1 | Bouras et al. | 12 sec |
| 2 | Jazayeriy et al. | 28 sec |
| 3 | Liu et al. | 27 sec |
| 4 | Proposed recommender | 2.6 milisec |

The experimental results and statistical evaluations demonstrate that the proposed Recommender approach provides promising results in terms of time efficiency and quality recommendations. The recommender system presented in this paper is more efficient and less time consuming in comparison to other existing recommenders.

**6. Conclusion and Future Work**

In this paper a novel CF recommendation approach is proposed which has the ability to handle heterogeneous data such as textual reviews, ranks, votes and views in a big data Hadoop environment with Cassandra database to guarantee the high response time to generate recommendations. In the proposed system, opinion based sentiment analysis is used to extract a hotel feature matrix and stored in a database. Our approach combines lexical analysis, syntax analysis and semantic analysis to understand sentiment towards hotel features. The NLTK library is used to identify polarity of the textual reviews. The system makes use of fuzzy rules to determine the hotel class depending upon the guest type. For accurate recommendations, Euclidean distance as an effective similarity measure is used. The proposed hotel recommendation system is a beneficial tool that recommends hotels to users according to their choices based on the type of guest (solo, family, couple etc). We have tested the performance gains of using Cassandra in terms of response time as it took 2.65 milliseconds to generate recommendations. It has shown significant performance to help reduce the system execution time by improving efficiency and scalability. The F-measure ratio has resulted in 0.950 approximately 95% which demonstrate that the proposed recommender approach provides promising results in terms of time and accuracy improvement which help consumers to get recommendations which are related to their choice and type which will impact their future behavior to use recommenders.

In future, we will study methods and techniques which will allow recommender systems to automatically use updated reviews and ratings online from the websites dynamically to give fresh recommendations. However, in operation, several new techniques should also be adopted by the operating websites in order to find out if the users have welcomed the resultant hotel recommendations. This can be done, for example, by integrating web cookies in order to capture user navigational traces and behaviors, and by obtaining users' feedbacks on new recommended items. New advancements are required to be researched that can drastically enhance the versatility of recommender frameworks.

**Conflict of Interest:** The authors declare that there is no conflict of interest regarding the publication of this paper.

**Data Availability Statement:** The authors will provide the data used for the experiments, if requested.

**Funding Statement:** No Funding is involved in this research.